\documentclass{aastex}
\usepackage{emulateapj5,epsfig}

\font\tenbg=cmmib10 at 10pt

\def \rvecmu{{\hbox{\tenbg\char'026}}}

\begin{document}

\title{Funnel Flows from Disks
to Magnetized Stars}

\author{A.V. Koldoba \altaffilmark{1},
R.V.E. Lovelace \altaffilmark{2},
G.V. Ustyugova \altaffilmark{3},
M.M. Romanova \altaffilmark{2}}

\altaffiltext{1}{Institute of
Mathematical Modelling,
Russian Academy of Sciences, Moscow,
Russia, 125047}
\altaffiltext{1}{Department of Astronomy,
Cornell University, Ithaca, NY 14853-6801;
rvl1@cornell.edu; romanova@astrosun.tn.cornell.edu}

\altaffiltext{3}{Keldysh Institute of
Applied Mathematics, Russian Academy
of Sciences, Moscow, Russia, 125047,
ustyugg@spp.Keldysh.ru}

\begin{abstract}

  This work considers flows from an 
accretion disk corotating with  the aligned
dipole magnetic  field of a rotating star.
  Ideal magnetohydrodynamics (MHD) is assumed
with the pressure and density related
as $p \propto \rho^\gamma$ and with
$\rho {\bf v}^2 \ll {\bf B}^2/4\pi$,
where ${\bf v}$ is the flow velocity.
 This limit corresponds to the Alfv\'en
radius for the disk
accretion larger than the corotation radius
and is of particular interest
because flow can be solved rigorously. 
     Transonic flows, which go from subsonic
motion near the disk to supersonic
inflow near the star,
are shown to be possible only for
a narrow range of $R_d \sim r_c \equiv
(GM/\Omega^2)^{1/3}$, where $R_d$
is the radius at which the dipole field
line intersects the disk, $r_c$ 
the corotation radius, $M$ the mass
of the star, and $\Omega$ its angular
rotation rate.  
    Over a larger range of
$R_d \lesssim r_c$, subsonic flows from the disk
to the star are possible.
  The transonic flows have very different
behaviors for $\gamma > 7/5$ and $< 7/5$.
    In both cases, the plasma flow velocity
${\bf v}$ (which is parallel to ${\bf B}$)
increases with decreasing distance $R$ from
the star.   
 However, for $\gamma >7/5$, the Mach number 
${\cal M} \equiv |{\bf v}|/c_s$ (with
$c_s$ the sound speed) initially increases
with $R$ decreasing from $R_d$, but for 
$R$ decreasing from $\approx 0.22 R_d$ (for $\gamma=5/3$)
the Mach number surprisingly {\it decreases}.
   In the other limit,
$\gamma < 7/5$, ${\cal M}$
increases monotonically with decreasing $R$.
   Application of these results is made
to funnel flows to rotating magnetized neutron stars
and young stellar objects.  
    The  spatial
variation of {\it both} the flow velocity 
and the Mach number are
crucial to the nature of the standing
shock wave near the star and to the
determination of emission line profiles.

\end {abstract}


\section{Introduction}

Models of magnetohydrodynamic (MHD) disk
accretion to a rotating star with
an aligned dipole magnetic field
have been discussed by many authors
(e.g., Ghosh \& Lamb 1979; Camenzind 1990;
 K\"onigl 1991;
Lovelace, Romanova, \&
Bisnovatyi-Kogan 1995).
   Major uncertainties remain however
because of the many assumptions required
to solve the MHD equations.  
  For example, the validity of the
standard formula for the disk's Alfv\'en
radius (Shapiro \& Teukolsky 1983) is
unknown and in fact this radius may
depend on the rotation rate of the star
(Lovelace, Romanova, \& Bisnovatyi-Kogan 1999).

     Definite predictions
can be made about funnel flows 
from  disks to magnetized stars
assuming a dipole magnetic field.
   These flows shown 
in Figure 1 naturally
explain the large infall
velocities observed  in many
T Tauri stars (Hartmann, Hewett,
\& Calvet 1994; 
Martin 1996; Najita {\it et al.} 2000). 
  Such flows may also occur in 
disk accretion to rotating magnetized
neutron stars (Li \& Wilson 1999). 
   Here, a study is made
of stationary, 
axisymmetric ideal 
magnetohydrodynamic (MHD) flows 
from an accretion  disk
to a rotating star with
an aligned dipole magnetic field.
  The MHD equations lead to a 
Bernoulli equation 
which we analyze in the general
case where pressure 
and density of a fluid particle 
are related as
$p \propto \rho^\gamma$ at constant
entropy.   
   A Bernoulli equation
for the isothermal case $p \propto \rho$
was discussed earlier
by Li and Wilson (1999).  
   Detailed calculations of emission line
shapes were made by
Hartmann {\it et al.} (1994)
and Martin (1996)
for a model where  plasma
free-falls along dipole field lines
of an aligned rotator, but these
authors do not analyze the
Bernoulli equation and do not
consider the sonic point of the flow.

\begin{thefigure} 
\epsscale{0.6}
\plotone{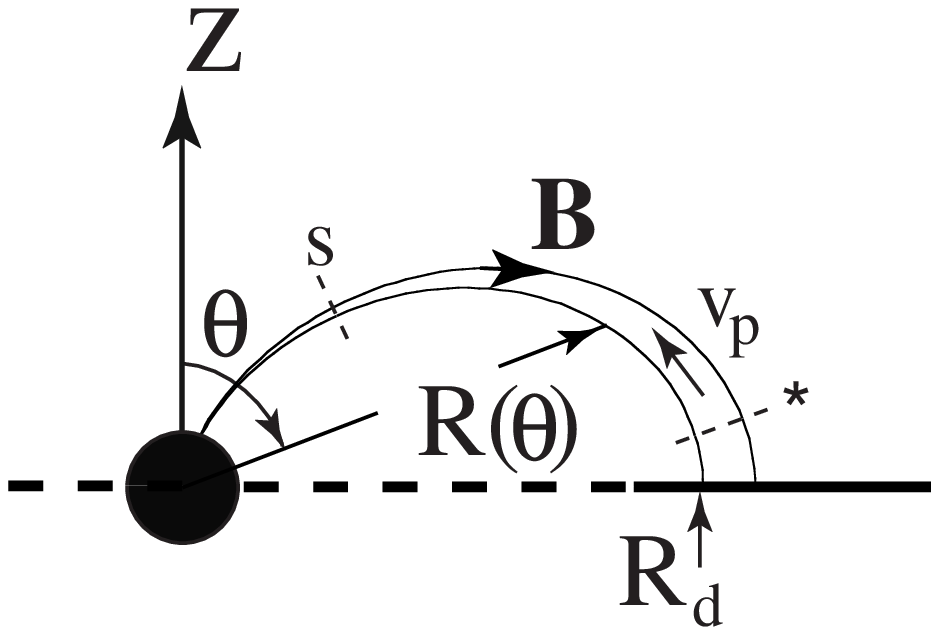}
\figcaption{
The thefigure shows a plasma 
accretion disk corotating 
with the aligned magnetic dipole field of a  star.
   Plasma  flows from the disk to the star
along a flux tube of the dipole, going from
subsonic  to supersonic flow.
    Conservation of magnetic flux gives $B_p A_p=$
const, and conservation of mass gives
$\rho v_p A_p=$ const, where $v_p$ is
the plasma speed along the field line and  $A_p$ is
the cross sectional area of the flux tube.
   The dashed line marked with (*)
indicates the sonic point and that marked with
(s) indicates
a standing shock.
 }
\label{Figure 1}
\end{thefigure}

    We use  inertial,
cylindrical $(r,\phi,z)$ and
spherical $(R,\theta,\phi)$
coordinate systems.
     As in the case of matter 
outflow along   open magnetic
field lines threading a disk
(Ustyugova {\it et al.} 1999), 
there are a number of integrals
of the plasma motion which
are conserved along magnetic 
field lines labeled by
the flux function $\Psi(r,z)=rA_\phi =$const,
where $A_\phi$ is the vector potential.
    The constant $K(\Psi)$ arises from
the conservation of mass, $\Omega(\Psi)$
from conservation of helicity,
$\Lambda(\Psi)$ from conservation of
angular momentum, $S(\Psi)$ from conservation
of specific entropy, and $E(\Psi)$ from
conservation of energy (Bernoulli's constant)
(see for example Ustyugova {\it et al.} 1999).
\begin{equation}
{\bf v}_p = \frac{K(\Psi)}{4 \pi \rho} 
{\bf B}_p~,\quad\quad
\Omega(\Psi)=\omega - 
\frac{KB_\phi}{4 \pi \rho r} ~,
\end{equation}
\begin{equation}
\Lambda(\Psi)=\omega r^2 - 
\frac{r B_\phi}{K} ~,\quad\quad\quad
S=S(\Psi)~,\quad
\end{equation}
\begin{equation}
 E(\Psi)=\frac{v_p^2}{2} + \frac{1}{2} 
(\omega - \Omega)^2r^2+w+\Phi_g-
\frac{\Omega^2 r^2}{2}~,
\end{equation}
where the $p$ subscript indicates
the poloidal $(r,z)$ [or $(R,\theta)$] part
of the vector,
$\omega = {v_\phi}/{r}$  the angular velocity of the
matter,
$w$ is the specific enthalpy, and $\Phi_g
=-GM/R$ the gravitational potential of
the star.  (The mass of the disk is assumed
negligible.)
   An
additional equation - the
Grad-Shafranov equation - follows 
from the equilibrium of forces
across magnetic field lines
 (Lovelace {\it et al.} 1986).

  In \S 2 we discuss  simplifications
which occur in the strong field limit,
and in \S 3 we analyze the Bernoulli 
equation (3).   In  \S 4 
dimensionless variables are introduced.
 In \S 5 we discuss the nature of
the transonic solutions which go from
subsonic flow near the disk to supersonic
flow near the star and show that the
behavior of the flows is different
for $\gamma < 7/5$ and $\gamma >7/5$.
In \S 6, we describe the nature of the
possible subsonic outflows from the
disk to the star.
  In \S 7,
we contrast the considered flows with 
spherical Bondi accretion. 
    In \S 8 we discuss conditions
for the strong field approximation
to hold. 
   In \S 9 we  apply 
the theory to accretion to a neutron 
star and to a young stellar object.
  In \S 10 we summarize the work.

\section{Strong Field Approximation}

   We  investigate the case 
where the magnetic field of the
star is dominant. 
    Namely, we suppose that the star has
dipole magnetic moment
$\rvecmu$ in the direction
 coinciding with  rotation axis of
the disk. 
    Thus the poloidal magnetic field 
in the considered
region is the dipole magnetic,
$
{\bf B}_p = 3  {\bf R}
({\rvecmu}\cdot{\bf R})/{R^5} -
{\rvecmu}/{R^3}$.
Further, we suppose that the
 magnetic field energy-density
is dominant compared with matter energy-density,
\begin{equation}
\frac{B_p^2}{8 \pi } \gg 
(~p~, ~~\rho v^2 ~,~~\rho \Phi~)~.
\end{equation}
Hence the
matter flow does not disturb  
the dipole magnetic field
significantly,
and the toroidal component of
the magnetic field is small,
$|B_\phi| \ll |{\bf B}_p|$.
Consequently, the full Grad-Shafranov
equation is not needed.   
   In \S 8
we give necessary conditions for the
strong field approximation to hold.

    In spherical coordinates, 
the flux function of the 
dipole magnetic field is
$
\Psi = \mu \sin^2( \theta)/{R}$.
Thus, the equation for the dipole field lines
is
\begin{equation}
R=R_d(\Psi) \sin^2 \theta,
\end{equation}
where $R_d(\Psi) = {\mu}/{\Psi}$ is the radius at
which the magnetic field line $\Psi=$const
passes through the disk ($z=0$).  
   Note that $B_R=(R^2\sin\theta)^{-1}
\partial \Psi/\partial\theta = 2\mu\cos\theta/R^3$,
and $B_\theta =-(R\sin\theta)^{-1}\partial \Psi/\partial R
=\mu \sin\theta/R^3$.
    Figure 1 shows the envisioned geometry.

    Equations (1) and (2) can be solved for
$\omega$ and $B_\phi$ to give
\begin{equation}
\omega = \Omega~ \frac{1-(\rho_A/\rho)
 (h /  r^2)}{1-\rho_A /
\rho}~,~~~
B_\phi = r \Omega \sqrt{4 \pi \rho_A}
\frac{1-h/r^2}{1-\rho_A / \rho}~,
\end{equation}
where $\rho_A \equiv {K^2}/(4 \pi)$ and
$h\equiv{\Lambda}/{\Omega}$.
   Note that  $\rho_A / \rho =
v_p^2/v_{Ap}^2$ is the poloidal Alv\'en
Mach number squared calculated 
using the  Alfv\'en
velocity $v_{Ap} \equiv 
|{\bf B}_p|/\sqrt{4\pi \rho}$.
   Owing to our assumptions,
$\rho_A/ \rho \ll 1$  
everywhere in the considered region.
  Consequently,
\begin{equation}
\omega - \Omega = \Omega \frac{\rho_A}{\rho} 
\left(1-\frac{h}{r^2}
\right)~,~~~
B_\phi = \Omega r \sqrt{4 \pi \rho_A}
\left( 1 - \frac{h}{r^2} \right)~.
\end{equation}
  We assume further that $(\rho_A/\rho)|1-h/r^2|
\ll 1$ with the result that 
$|\omega - \Omega | \ll \Omega.$
From equation (13) note that $|B_\phi|/|{\bf B}_p|
=(r\Omega/v_{Ap})|1-h/r^2| \ll 1$.

   Thus a given fluid particle
moves along the poloidal
magnetic field ${\bf v}_p \propto {\bf B}_p$,
and its angular velocity $\Omega(\Psi)$ is 
a constant.  
  The value $\Omega (\Psi)$ is
the angular velocity
of rotation of the footpoint of the magnetic field
line $\Psi=$const, which we suppose is frozen
into the star. 
   Thus, $\Omega (\Psi) = \Omega_{star} =$const is
the angular rotation rate of the star.
  We omit the subscript  on $\Omega$ in
the following.

\section{Bernoulli's Constant}

    Consider now the matter flow along 
a specific magnetic field
line $R=R_d(\Psi) \sin^2 \theta$. 
    The effective
potential along this line in a reference
frame rotating with rate $\Omega$
is
\begin{equation}
\Phi (R) = \Phi_g + \Phi_c = -\frac{GM}{R} -
\frac{\Omega^2 R^2 \sin^2 \theta}{2} = - \frac{GM}{R} -
\frac{\Omega^2 R^3}{2 R_d}~.
\end{equation}
    The magnitude of the magnetic 
field along this field line is
\begin{equation}
B_p(R) = \frac{\mu}{R^3}(4-3\sin^2\theta )^{1/2}=
\frac{\mu}{R^3} \left(4-\frac{3R}{R_d}\right)^{1/2}~.
\end{equation}
Thus the Bernoulli constant along this field line is
\begin{equation}
E=F(R,\rho)=\frac{K^2 \mu^2}{32 \pi^2 \rho^2 R^6}
\left( 4 - \frac{3R}{R_d} \right) 
+ w +\Phi(R)~,
\end{equation}
where $w =S \rho^{\gamma -1}/(\gamma -1) 
= c_s^2/(\gamma -1)$, with
$c_s$ the sound speed and 
$S \equiv \gamma p/\rho^\gamma$
the specific entropy.

    We  consider flows which are
subsonic near the disk, $|v_p|
\ll c_s$ for  $R \lesssim R_d$.
  There are then three possible
cases:  (1) The flow is transonic, going
from subsonic to supersonic at some
distances $R_* < R_d$;  (2) The flow remains
subsonic between the disk and the star;
or (3) no stationary solution to the
Bernoulli equation exists.

   For the transonic flows, 
note that the $\rho$ derivative of equation
(10) is
\begin{equation}
\frac{\partial F}{\partial \rho}\bigg|_R 
= \frac{c_s^2-v_p^2}{\rho}~,
\end{equation}
in that
$\left( {\partial w}/{\partial \rho} \right)_S =
{c_s^2}/{\rho}$.
   In general $F(R,\rho)$ has a minimum as
a function of $\rho$ at, say, $\rho_m(R)$,
and $E=F(R,\rho)$ has two solutions for $\rho$.
 The larger  $\rho $ solution
has
$\partial F / \partial \rho > 0 $
and corresponds to subsonic flow. 
The smaller $\rho$ solution  has
$ \partial F / \partial \rho < 0 $ and
corresponds to supersonic flow.

   For the transonic flows,
the conditions for a smooth 
transition 
through the sonic point are
\begin{equation}
 \frac{\partial F}{\partial R} \bigg|_* =0~,
\quad \quad
 \frac{\partial F}{\partial \rho} \bigg|_* =0~.
\end{equation}
  In order to pass from subsonic to
supersonic flow, this extremum of $F(R,\rho)$
must be a saddle point which corresponds to
$F_{RR*}F_{\rho \rho*}-F_{\rho R*}^2
<0$, where the subscripts indicate
partial derivatives.  
    The conditions for this are
given in \S 5.

 \section{Dimensionless Variables}

   With $R$ measured in units of $R_d$,
we can write the effective potential as
$$
\Phi = -\Omega^2 r_c^2\left( {\alpha \over R/R_d} + 
{(R/R_d)^3 \over 2\alpha^2} \right)~,
$$
where $\alpha \equiv r_c/R_d$, and the 
characteristic length 
\begin{equation}
r_c \equiv
\left({GM \over\Omega^2}\right)^{1/3} 
\approx 1.5\times 10^8{\rm cm}\left({M \over M_\odot}
\right)^{1/2}\left({P \over 1{\rm s}}\right)^{2/3}
\end{equation}
  is the 
``centrifugal radius''  where
the centrifugal force 
balances the gravitational
attraction of the star ( in the 
$z=0$ plane in
the absence of magnetic forces),
and $P=2\pi/\Omega$ is the star's
rotation period.  
    For a young star of solar mass,
$r_c\approx 2.9\times 10^{11}
{\rm cm}(P/{\rm 1d})^{2/3}$.

Useful dimensionless variables are
\begin{eqnarray}
\hat{R}=\frac{R}{R_d}~,\quad \quad \quad
\hat{\rho}= \frac{4 \pi \alpha 
\Omega R_d^4}{K \mu} ~\rho~, 
\quad \quad \quad \quad ~~~ \nonumber\\
{\hat{E}} = \frac{E}{ \Omega^2 r_c^2}~,\quad~~~
\hat{S} = \frac{S}{\Omega^2 r_c^2}
\left( \frac{K \mu}{4 \pi \alpha \Omega R_d^4}
\right)^{\gamma -1}~,~
\end{eqnarray}
and $\hat{F}=F/(\Omega r_c)^2$, $(\hat{c}_s)^2=
\hat{S}\hat{\rho}^{\gamma-1} = c_s^2/(\Omega r_c)^2$,
$\hat{v}_p = v_p/(\Omega
r_c)=(4-3\hat{R})^{1/2}/(\hat{\rho}\hat{R}^3)$, where
$\alpha=r_c/R_d$.  Note that at $\hat{R}=1$,
$\hat{\rho}\hat{v}_p =1$. 
    For simplicity of notation we now drop the hats.
  We then have $ R_{star}/R_d < R\leq 1$, where
$R_{star}$ is the star's radius.

    In terms of these
variables, Bernoulli's equation  becomes
\begin{equation}
E=F(R,\rho)=\frac{4-3{R}}{2\rho^2{R}^6} + \frac{S
\rho^{\gamma-1}}{\gamma-1} -
\left( \frac{\alpha}{{R}} + 
\frac{{R}^3}{2 \alpha^2}
\right) ~.
\end{equation}
    We suppose  that the disk 
matter is at a relatively low temperature, that is,
$w(R_d)=c_s^2(R_d)/(\gamma-1) \ll - \Phi (R_d)$,
 and that the
outflow speed from the disk is small
$v_p^2(R_d) \ll - \Phi (R_d)$. 
   Under these conditions the dimensionless 
quantities at the
starting point of the flow ($R=1$) are:
the Bernoulli constant,
$E \approx -  \alpha - 1/2\alpha^2$,
the sound speed,
$c_s(1) \ll 1$, and the density,
$\rho(1) \gg 1$.  Because 
$S=c_s^2/\rho^{\gamma-1}$, we also
have $S=S(1) \ll 1$.

\section{Transonic Flows}

    The mentioned conditions 
for a smooth transition 
through the sonic point give
\begin{eqnarray}
0&= &\frac{\partial F}{\partial {R}}\bigg|_\star =
-\frac{3(8-5{R_*})}{2\rho_*^2{R_*}^7} +
 \frac{\alpha}{{R_*}^2} -
\frac{3{R_*}^2}{2\alpha^2}
 ~, \\
0&=& \frac{\partial F}{\partial \rho}\bigg|_\star = -
\frac{4-3{R_*}}{\rho_*^3{R_*}^6} + S
\rho_*^{\gamma-2}~.
\end{eqnarray}
  In general, 
$F_{\rho\rho*} >0$.
  In order for this critical point to be
a saddle point of $F(R,\rho)$, we must
have $F_{RR*} <F_{\rho R*}^2/F_{\rho\rho*}$.
  This requirement is satisfied if
$R_* > (N/D)^{1/4} (2\alpha^3/3)^{1/4}$,
where $N=30\gamma R_*^2-45R_*^2+140R_*-100\gamma R_*
+80\gamma-112$ and
$D=60\gamma R_*^2-15R_*^2+52R_*-
188\gamma R_*+144\gamma-48$.
  This condition on $R_*$ is always
satisfied if $\gamma \leq 7/4$.
  For $\gamma=5/3$, the condition
is satisfied for $R_* >0.581
(2\alpha^3/3)^{1/4}$.
  The  nature of the function $F(R,\rho)$
can be understood from its contour plot shown
in Figure 2a for a sample case.

\begin{thefigure} 
\epsscale{1.}
\plotone{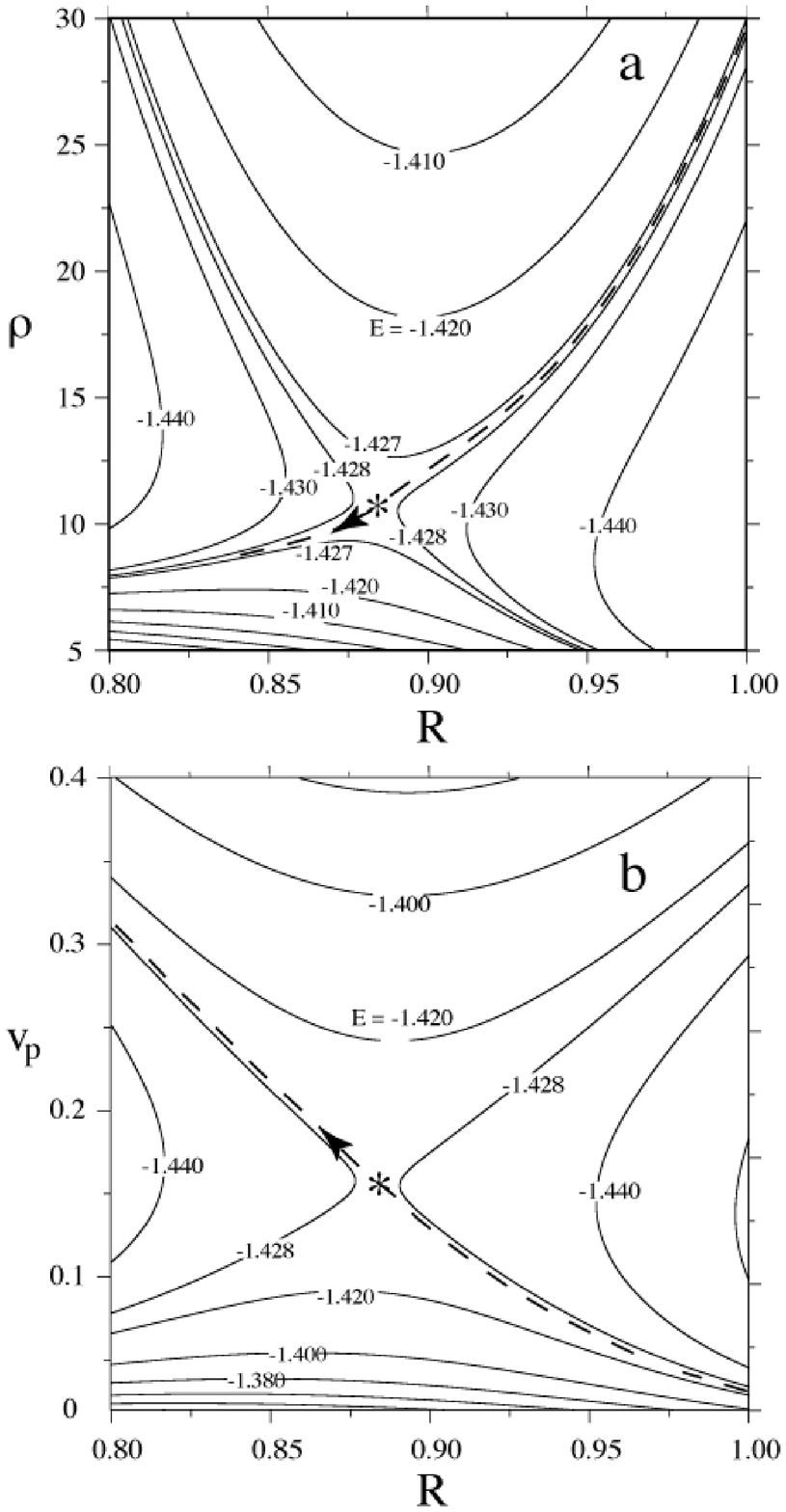}
\figcaption{
Contours 
of the Bernoulli function $E=F(R,\rho)$ in
the top panel $({\bf a})$ and $E=F(R,v_p)$
in the bottom panel $({\bf b})$, for 
a sample case with $\gamma=5/3$,
dimensionless entropy $S=0.005$, and $\alpha=1$.
  Distance from the center
of the star $R$ is
normalized by $R_d$.
   The transonic flow starts at $R=1$ and
follows the dashed line.
   The sonic  point where $v_p=c_s$ - indicated by
the asterisk - is  the saddle point
of the surface $F$ at $(R_*, \rho_*)$.
   For $R>R_*$ the
flow is subsonic and $\partial F/\partial \rho >0$,
while for the reversed inequalities the flow is
supersonic.
  A possible subsonic flow would follow the
upper $E=-1.42$ contour in $({\bf a})$ and
the lower $E=-1.42$ contour in $({\bf b})$.
}
\label{}
\end{thefigure}

  The term in equation (16) involving $\rho_*$
can be rewritten in terms of $S$ using 
equation (17).  
  This term is then seen to be negligible
compared with the terms involving $\alpha$
because $S \ll 1$.
   Equation (16) is then seen to
correspond to a balance of
the gravitational attraction
and centrifugal force along
the field line. 
    Under this condition,
equation (16) can be satisfied only for
\begin{equation}
\alpha \leq \alpha_0 \equiv (3/2)^{1/3} \approx
1.145~.
\end{equation}
Consequently,
$$
 R_*\approx\left({2 \alpha^3 \over 3}\right)^{1/4}~,
\quad
\rho_*\approx S^{-{1\over \gamma+1}}\left( {4-3R_* \over  
 R_*^6 }\right)^{1 \over\gamma+1}~, 
$$
\begin{equation}
c_{s*}=\left(S\rho_*^{\gamma-1} \right)^{1/2} \approx
S^{1\over \gamma+1}
\left({4-3R_* \over R_*^6} \right)^{\gamma-1 \over
2(\gamma+1)}~.
\end{equation}
Thus the mass flux-density is
$\rho_* v_{p*} =\rho_* c_{s*}=(4-3R_*)^{1/2}/R_*^3$.
   The approximation involves neglecting the
$\rho_*$ term in equation (16) which is
valid for $S \ll 1$ and $\alpha$ not too
small compared with unity.   
   In the next approximation, we find 
$R_*\approx(2\alpha^3/3)^{1/4}
[1-1.1S^{3/4}\alpha^{-11/8}]$ for $\gamma=5/3$,
and $R_*\approx(2\alpha^3/3)^{1/4}
[1-0.88S^{0.87}\alpha^{-0.84}]$ for $\gamma=1.3$,
and both expressions for $R_* \ll 1$.
    If the factor in the square brackets is
less than about $0.863$ then as mentioned
 below equation (17) the saddle point disappears
(by becoming a minimum).
  The saddle point given by equation (19)
is the only saddle point with $R_*\leq 1$.
   Figure 3 shows the  dependences of $R_*$ and
$c_{s*}$ obtained by solving equations (16) and
(17) numerically without approximations.

\begin{thefigure} 
\epsscale{0.6}
\plotone{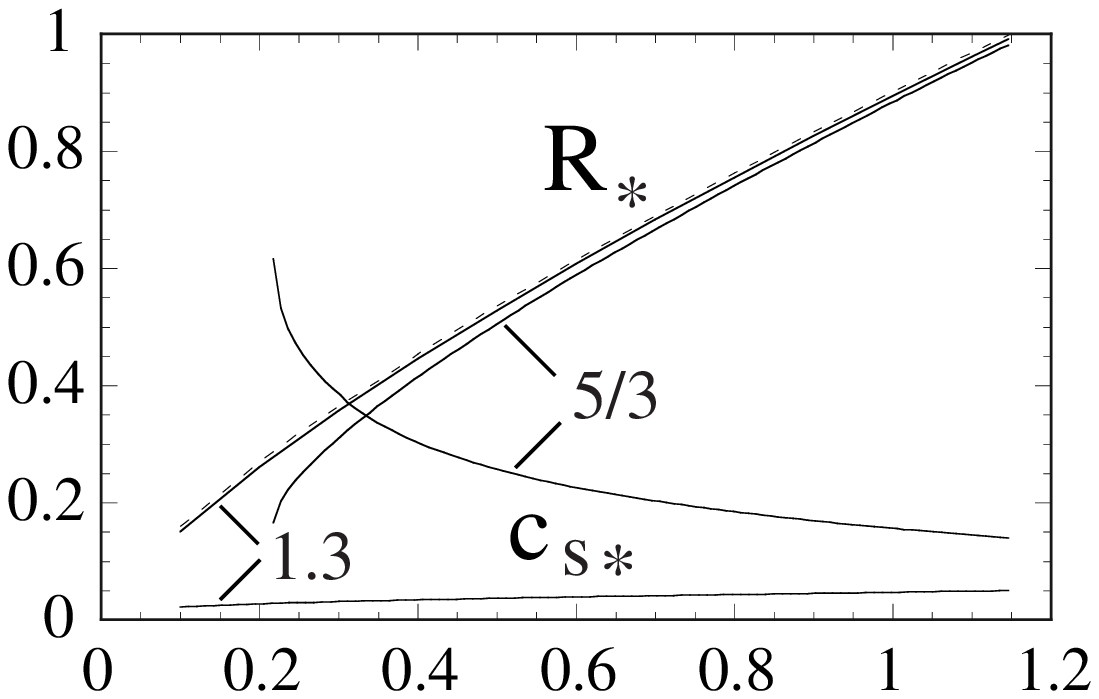}
\figcaption{
Dependences of 
the sonic point distance
$R_*$ and  sound speed $c_*$ on $\alpha = r_c/R_d$
for the cases of $\gamma=5/3$ and $\gamma=1.3$
and dimensionless entropy $S=0.005$.
  The dashed line is for the
approximation of equation (19). 
  For the $\gamma=5/3$ case, the saddle point
of $F(R,\rho)$ changes into  a minimum
 for $\alpha <0.218$ so that
these values are not of interest.
}
\label{Figure 3}
\end{thefigure}

    Figures (4) and (5) show sample radial
profiles of the fluid variables for 
$\gamma=5/3$ and $1.3$, respectively.
  A value of $\gamma$ smaller than the
ideal monotonic gas value is of interest
for  a high temperature plasmas
where the  electron heat conduction
can be modeled by smaller $\gamma$
(as is the case for the solar wind).
   The difference in behavior of
the cases shown in Figures (4) and
(5) can be understood by considering
the Bernoulli equation for $R\ll 1$.
    In this limit $v_p \approx (2\alpha/R)^{1/2}$
[in dimensional variables this is $v_p
\approx (2GM/R)^{1/2}$, which is
the free-fall speed], $\rho \approx
(2/\alpha)^{1/2} R^{-5/2}$, and 
$c_s \approx (2/\alpha)^{(\gamma-1)/4}
S^{1/2} R^{-5(\gamma-1)/4}$.
   Hence the Mach number 
varies as ${\cal M} = v_p/c_s \approx
2(\alpha/2)^{(\gamma+1)/4} S^{-1/2}
R^{(5\gamma-7)/4}$.  
    Thus for $\gamma >7/5$, the
Mach number initially increases
with decreasing $R$, but for
small enough $R$, ${\cal M}$
decreases with decreasing $R$. 
   The radius of the maximum
of ${\cal M}$ is found
to be $R_{max} \approx (5\gamma-7)
/\{6(\gamma-1)[1+1/(2\alpha^3)]\}$.
  The maximum value of ${\cal M}$
is therefore $\propto S^{-1/2}$.
   On the other hand for $\gamma <7/5$,
${\cal M}$ increases monotonically
with decreasing $R$.
    Therefore, the changeover in
behavior between Figures (4) and
(5) occurs at $\gamma=7/5$.
  The radial variation of both
$v_p$ and ${\cal M}$ are clearly
important for calculations of
the emission line profiles
(Hartmann {\it et al.} 1994).

   For $\gamma>7/5$, the Mach
number does not decrease
through ${\cal M}=1$.
  For $R \ll 1$, Bernoulli's equation
is $2/(\rho^2R^6)+S\rho^{\gamma-1}/(\gamma-1)
\approx \alpha/R$.
   Substituting $\rho = \eta/R^p$
with $p=6/(\gamma+1)$ gives
$2/\eta^2 +S \eta^{\gamma-1}/(\gamma-1) \equiv
L(\eta) \approx \alpha R^{(5\gamma-7)/(\gamma-1)}$.
   Clearly, $L(\eta)$ has a minimum
value of $L_{min}=
[2+4/(\gamma-1)](S/4)^{2/(\gamma+1)}$
at $\eta_{min}=(4/S)^{1/(\gamma+1)}$
which corresponds
to ${\cal M}=1$.
   Thus for $\gamma >7/5$, there is
a minimum radius where a stationary
flow exists, $R_{min} =
[(2/\alpha)(\gamma+1)/(\gamma-1)]^{(\gamma+1)/(5\gamma-7)}
(S/4)^{2/(5\gamma-7)}$.
   For most conditions this
radius is expected to be less than the
radius of the star.

    Near the star the flow is expected to
have a standing shock where ${\cal M}$
discontinuously decreases to
a value less than unity. 
   For $\gamma<7/5$ there
also must be a shock.
   Note that the ratio of the
temperatures across this
shock for an ideal gas  is $T_2/T_1 =
1+2(\gamma-1)(\gamma{\cal M}^2+1)({\cal M}^2-1)
/[(\gamma+1)^2 {\cal M}^2]$ ($ \approx
2\gamma(\gamma-1){\cal M}^2/(\gamma+1)^2$ 
for ${\cal M} \gg 1$), where
$T_1$ is the upstream temperature and
$T_2$ is downstream (closer to the star).
   Calculation of the location
of this shock is beyond
the scope of this work.

\begin{thefigure} 
\epsscale{0.6}
\plotone{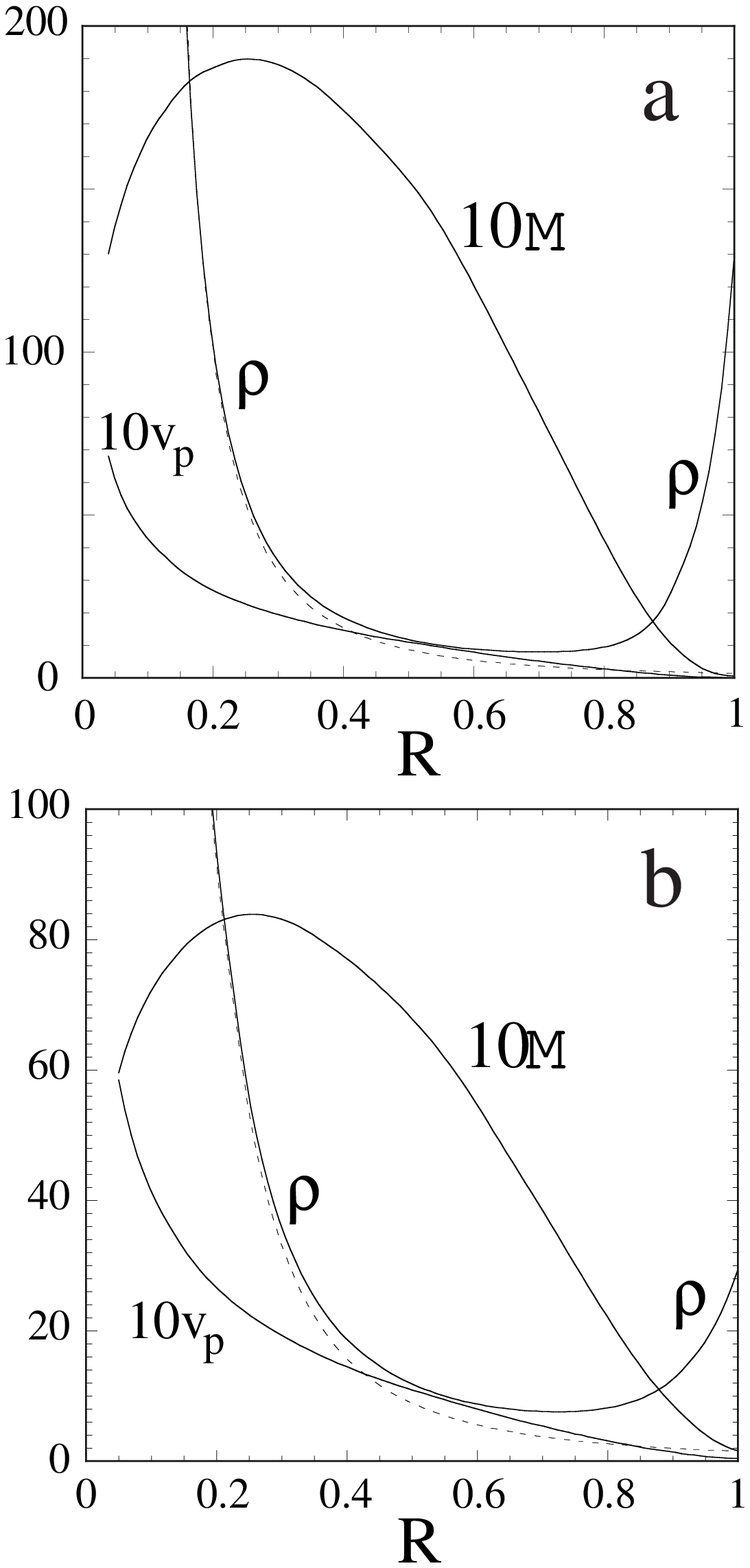}
\figcaption{
The thefigure shows the radial
variations of the dimensionless density $\rho$,
poloidal velocity $v_p$, and Mach number ${\cal M}=v_p/c_s$
for cases with $\gamma=5/3$, $\alpha=1$,
 and  dimensionless 
entropies $S=0.001$ (top panel, {\bf a}) 
and $S=0.005$ (bottom panel, {\bf b}).
  For (${\bf a}$), $E\approx-1.462$, $c_{sd} \approx 0.150$,
$\rho_d \approx 129.4$, $R_*\approx 0.898,$ and
$c_{s*}\approx 0.0841$.
  For (${\bf b}$), $E \approx-1.428$, 
$c_{sd} \approx 0.218$,
 $\rho_d \approx 29.4$, $R_* \approx 0.884$, and
$c_{s*}\approx 0.156$.
   The dashed curves represent const$R^{-5/2}$
fitted to a point at the top of the plot.
}
\label{Figure 4}
\end{thefigure}

\begin{thefigure} 
\epsscale{0.6}
\plotone{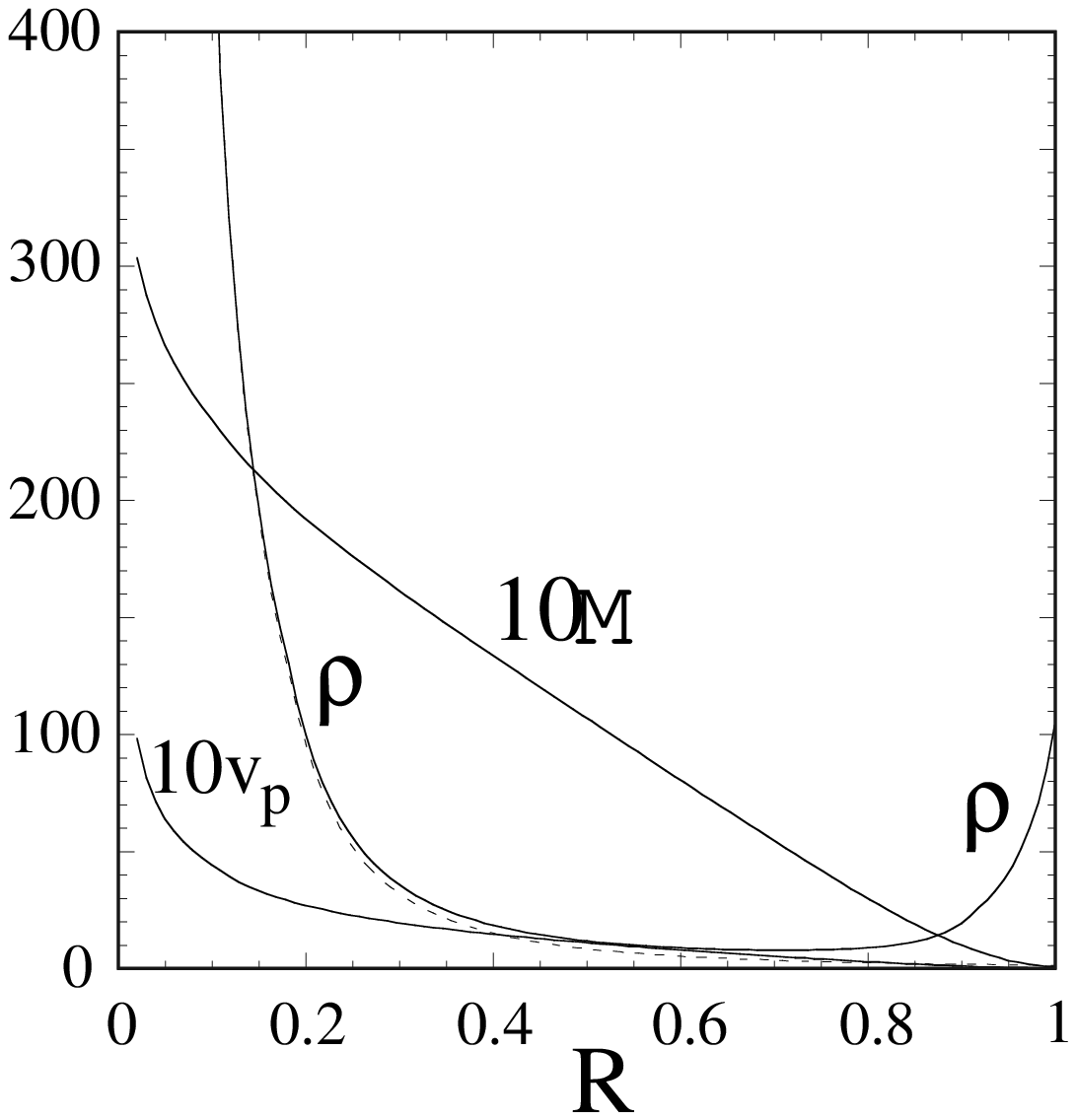}
\figcaption{
Radial
variations of the dimensionless density $\rho$,
poloidal velocity $v_p$, and Mach number ${\cal M}=v_p/c_s$
for a case with $\gamma=1.3$, $\alpha=1$,
and  dimensionless 
entropy $S=0.005$.
    The dashed curve represents const$R^{-5/2}$
fitted to a point at the top of the plot.
}
\label{Figure 5}
\end{thefigure}

\begin{thefigure} 
\epsscale{0.6}
\plotone{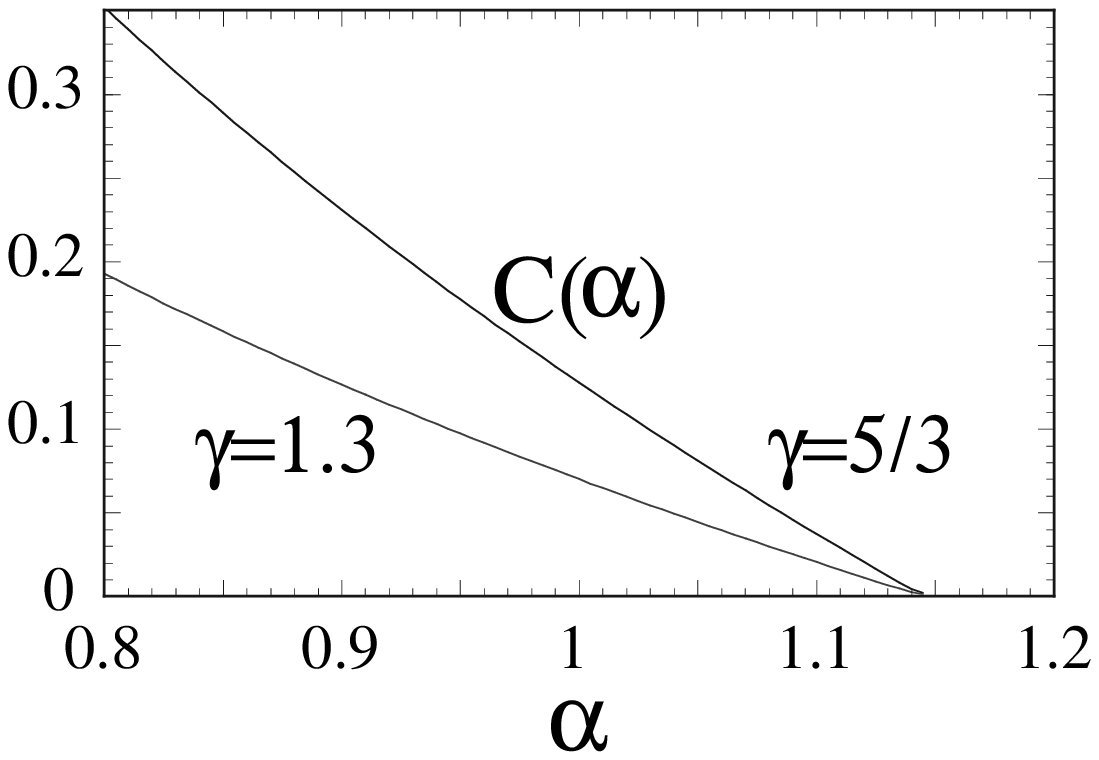}
\figcaption{
Dimensionless function in equation (20).
The dimensionless sound speed $c_{sd}$  must
be less than $C(\alpha)$.
}
\label{Figure 6}
\end{thefigure}

    The value of entropy $S$ for
transonic flows $S_{ts}$ is determined by
conditions at origin of the flow at
the disk where $v_{pd}^2 \ll c_{sd}^2$,
where $c_{sd}$ is the sound speed at the
 disk.
  At the disk,
$$
E={c_{sd}^2 \over \gamma-1}
-\left(\alpha +{1 \over 2 \alpha^2}\right)~,
$$
while at the sonic point,
$$
E={1\over 2}{\gamma+1 \over \gamma-1}
c_{s*}^2-\left({\alpha \over R_*} +
{R_*^3 \over 2\alpha^2}\right)~.
$$
Note that $c_{s*}^2= S\rho_*^{\gamma-1}$
and $\rho_*=(4-3R_*)^{1/2}/(c_{s*} R_*^3)$.
  Therefore,
in order to have a transonic flow, we need
\begin{equation}
S\!=\!S_{ts} \!\equiv \!
\left({2 \over \gamma+1}\right)^{\gamma+1
\over 2}
\left({R_*^6 \over 4-3R_*}\right)^{\gamma-1 \over 2}
\left[c_{sd}^2 -C^2(\alpha)\right]^{\gamma+1 \over 2}~,
\end{equation}
where 
$$
C^2(\alpha)=(\gamma-1)\left(\alpha +{1\over 2 \alpha^2} -c_1
\alpha^{1/4}\right)~
$$
is obtained by using equation (19), and
$c_1 = (3/2)^{1/4}+(2/3)^{3/4}/2 \approx 1.476$.
 Figure 6 shows $C(\alpha)$. 
  Clearly, for a given value
of $\alpha$ it is necessary to have
sufficiently high disk temperature or
disk sound speed, $c_{sd} \geq C(\alpha)$.
   This inequality and inequality (18)
for transonic flow can be satisfied only for
\begin{equation}
\alpha_{min} \leq \alpha \leq \alpha_0~,
\end{equation}
where $\alpha_{min}$ is such that
$c_{sd}=C(\alpha_{min})$.
  Inequalities (21) correspond to a 
limited range of $R_d$,
$ r_c/\alpha_0  <R_d < r_c/\alpha_{min}$.
   For example, for constant 
$c_{sd}=0.1$ and $\gamma=5/3$, we 
have $c_{sd}\geq C(\alpha)$ for $1.03 \leq \alpha
\leq 1.145$ and $S$ varies between $0$ and $0.00147$;
 for $c_{sd}=0.2$, the
allowed range is $0.928\leq \alpha \leq 1.145$
and $S$ varies between $0$ and $0.00933$.

\section{Subsonic Flows}

  For $\alpha_{min} \leq \alpha \leq \alpha_0$
(with $\alpha_{min}$ given above),
it is easy to show that there are outflows
from the disk to the star which are
subsonic.  
  Figure 7a shows the $R$ dependences
for such a case.
   For $R\ll 1$, equation (15) gives
$c_s \approx [\alpha(\gamma-1)/R]^{1/2}$ 
and $\rho \approx [\alpha(\gamma-1)/(S R)]^{1/(\gamma-1)}$.

  For $\alpha < \alpha_{min}$, 
there are {\it no}
stationary solutions 
to the Bernoulli equation (15).
  Equivalently, there are no stationary
outflows for $R_d > r_c/\alpha_{min}$.

  For $\alpha > \alpha_0$ or 
$R_d < r_c/\alpha_0$, there are  also
subsonic outflows from the disk to the star
as shown in Figure 7b.

\begin{thefigure} 
\epsscale{0.6}
\plotone{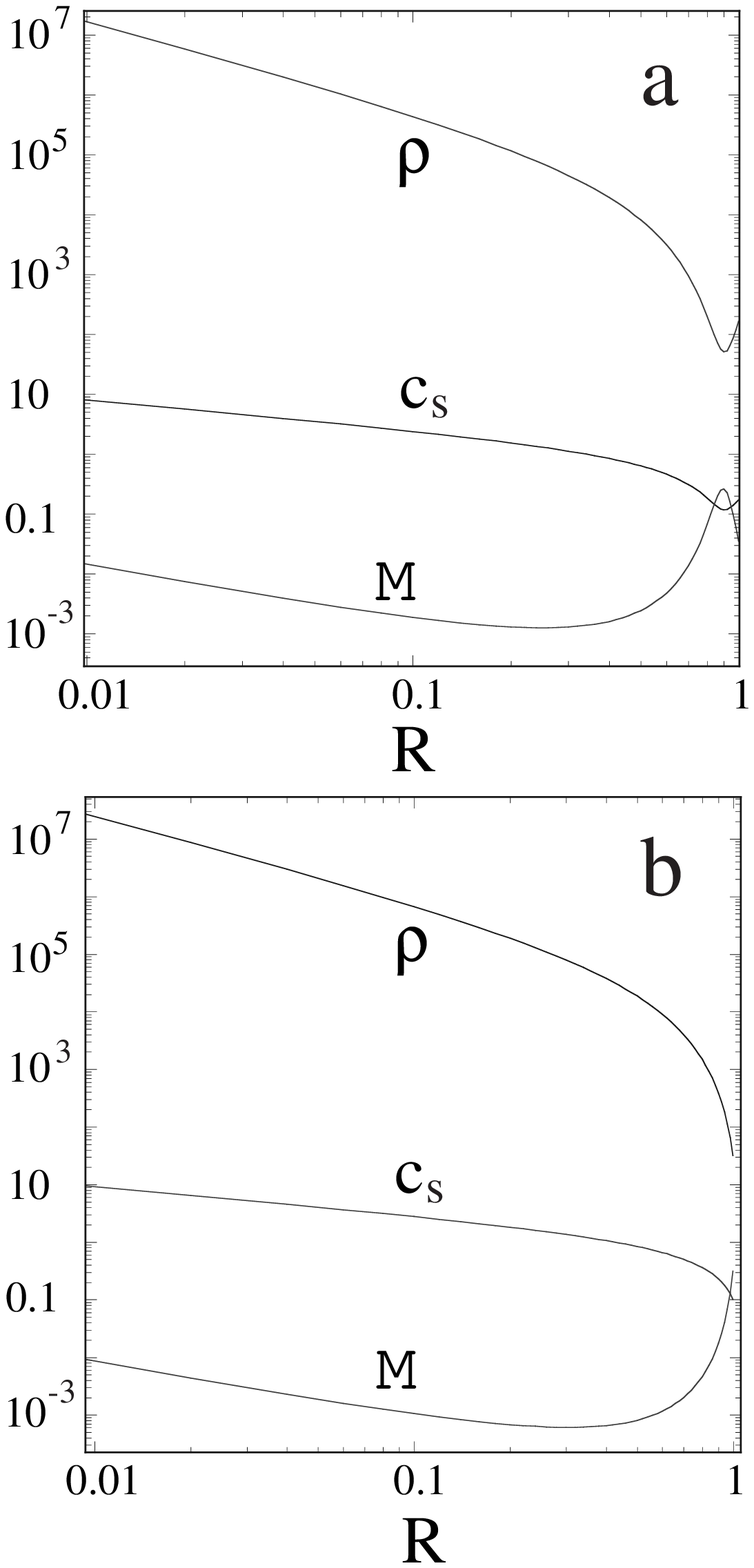}
\figcaption{
Radial dependences of density
$\rho$, sound speed $c_s$, and
Mach number for subsonic flows
from  disk to  star.
  The top panel $({\bf a})$ has $\alpha=1.0$,
$S=0.001$,  $\gamma=5/3$, and $c_{sd}\approx 0.173$. 
   In this
case the Mach number
rises to  about $0.26$ at $R=0.9$.
  The bottom panel $({\bf b})$ has $\alpha=1.3$,
$S=0.001$,  $\gamma=5/3$, and $c_{sd} \approx 0.1$.
}
\label{Figure 7}
\end{thefigure}

\section{Contrast with  Bondi Accretion}

 The Bernoulli equation(10) differs
from that for spherical
Bondi (1952) accretion in that here
the plasma flows along
the magnetic field lines,  the matter is
rotating, and the flow starts from a
finite radius of a
disk.
   For the Bondi flow,
mass conservation gives
$4\pi R^2 \rho v_R= \dot{M}=$const so that
the Bernoulli equation is 
\begin{equation}
E={ (\dot{M}/4\pi)^2 \over 2\rho^2 R^4}
+{S \rho^{\gamma-1} \over \gamma-1}
-{GM \over R}~,
\end{equation}
where $S=c_{s\infty}^2/\rho_\infty^{\gamma-1}$,
with
$c_{s\infty}$ and $\rho_\infty$  the
sound speed and density at a large distance
from the star.
  In contrast, the first term of equation (10)
is $\propto 1/(\rho^2 R^6)$ owing to
the ``focusing'' of the magnetic field, and
further, equation (10) includes the centrifugal
potential $-\Omega^2r^2/2$.
     For small $R$, the Bondi
flow has $v_R \approx (2GM/R)^{1/2}$
as in the for the up and over flow; however,
$\rho \sim 1/R^{3/2}$
and $c_s \sim 1/R^{3(\gamma-1)/4}$, whereas
for the funnel flow $\rho \sim 1/R^{5/2}$
and $c_s \sim 1/R^{5(\gamma-1)/4}$.  Thus,
the Mach number for the Bondi flow 
 ${\cal M} \sim 1/R^{(5-3\gamma)/4}$ always
increases (or is constant) with decreasing 
$R$ for $\gamma \leq 5/3$.

\section{Conditions for Strong Field Approximation}

  In the strong field approximation of \S 2,
the disk plasma is assumed to corotate with the star even 
though $r$ is less than the corotation radius $r_c$.
   Equilibrium of this part of the disk thus requires
an additional outward radial force which
arises naturally from
the slight bending of the field lines passing
through the disk.
   This radial magnetic force (per unit
area of the disk) is simply $F_r^{mag}=
-(B_rB_z)_h/2\pi>0$, where the $h-$subscript
indicates evaluation at the disk 
surface $z=h$  [equation (5) 
of Lovelace {\it et al.} 1995].
   Because $(r_c-r)/r_c \ll 1$, the required
radial magnetic field is 
small, $(B_r/B_z)_h=
(6\pi \sigma \Omega^2  /B_z^2)(r_c-r) 
=3(\Omega r_c/v_A)^2 h(r_c-r)/r_c^2 \ll 1,$
where $v_A$ is the Alfv\'en velocity
based on the midplane density of the disk
and $B_z(r,0)$.

   Also, in the approximation of \S 2 
the toroidal magnetic field is neglected.     
  A small deviation of the disk
rotation rate $\omega_d(r)$ from the star's
rate $\Omega$ can lead to a significant
toroidal magnetic at the disk,
$(B_\phi)_h = -(hr/\eta)(\omega_d-\Omega)B_z$,
where $\eta$ is the magnetic diffusivity of
the disk [equation (2) of Lovelace {\it et al.}
(1995)].
    If $\eta$ is of the order of the viscosity $\nu$ 
given by the
Shakura and Sunyaev (1973) prescription
$\nu =\alpha c_s h$ (with $\alpha <1$),
then, in order to have $|(B_\phi)_h/B_z|
\ll 1$, we must have $|r(\omega_d-\Omega)|
\ll \alpha c_s$.

\section{Applications}

    Table 1 summarizes the relevant parameters
for the cases of disk accretion  to 
(1) a rotating magnetized
neutron star, and to 
(2)  a rotating young stellar object.
  In both cases, the  
Alfv\'en radius $r_A$ 
(Shapiro \& Teukolsky 1983) is 
larger than $r_c$ so that the disk plasma is
expected to corotate with the star.
   The disk plasma may move radially across
the magnetic field by the interchange instability
(Kaisig, Tajima, \& Lovelace 1991; Rast\"atter \&
Schindler 1999).
    For the neutron star the surface magnetic
field ($\mu/R_{star}^3$) is assumed to be 
$10^{12}{\rm G}$, while for the
young star it is $10^4{\rm G}$.

  For the neutron star case, 
we assume $c_{sd}/(\Omega r_c)
=0.1$ (with dimensions restored), which corresponds
to a surface temperature of the disk $T_d \approx
4.1\times 10^7$K for a fully
ionized hydrogen plasma and $\gamma=1.3$
with mean particle mass $m_H/2$.
   The half-thickness of the disk is
$h \sim c_s/\Omega \approx 0.1r_c$.
   From  Figure 6 with $\gamma=1.3$,
 we find that transonic outflows from
the disk to the star
are possible for $\alpha$
in the range 
$\alpha = r_c/R_d \approx 0.95 - 1.145$.
which corresponds to 
$R_d \approx (0.873 -1.053)r_c$.
  The sonic point location 
$R_*/R_d=\sin^2(\theta_*)$ has 
$\theta_*$ varying between $69^\circ$
(the lower limit of $\alpha$ and
the upper limit of $R_d$) and
$90^\circ$ where the sonic point
is inside the disk.
   The height of the sonic point
$z_*=R_d \sin^2(\theta_*)\cos(\theta_*)$
is $\lesssim  0.273r_c$.
    Thus, transonic outflows
occur over an interval 
$\Delta R_d \approx 0.18 r_c$.
  In  dimensionless
variables, $\rho_d v_{pd}=1$.
   Therefore, if all of the disk accretion
goes into transonic flow along  the field lines,
$\dot{M}_{acc}= 2(2\pi)R_d\Delta R_d
K \mu/(4\pi R_d^3)$, where one of the
factors of two in the numerator accounts
for the two sides of the disk.
  Thus the characteristic density normalizing
$\rho$ in equation (14) is
\begin{equation}
\rho_0 ={K\mu \over 4\pi\alpha \Omega R_d^4}
={\dot{M}_{acc} \over 4\pi \alpha 
\Omega R_d^2 \Delta R_d}~.
\end{equation}
For the mentioned values and those in 
Table 1, we find  $\rho_0 \approx 1.4\times 10^{-9}
{\rm g}/{\rm cm}^3$.
   The ratio of the maximum
to minimum radii of 
the flow is $r_c/R_{star}=150$
 and that the free-fall speed at
the star is
$v_{ff}=(GM/R_{star})^{1/2}
\approx 1.1\times 10^{10}{\rm cm/s}$.
   The angle at which the flow hits
the star is $\theta_{star}=
\sin^{-1}[(R_{star}/r_c)^{1/2}] 
\approx 4.7^\circ$.
  At this point the flow velocity 
is at an angle $\approx \theta_{star}/2$
with respect to $\hat{{\bf R}}$.

  For the case of a young stellar object,
we assume $c_{sd}/(\Omega r_c) =0.05$,
which corresponds to a surface temperature
of the disk $T_d \approx 2,800$K for a
slightly ionized
hydrogen plasma and $\gamma=5/3$.
   The half-thickness of the disk is
$h \sim c_s/\Omega \approx 0.05r_c$.
  The height of the sonic point
is $z_* \lesssim 0.174$.
  The considerations for this case
are similar to those for the neutron
star.  The range of $\alpha$ is
$\approx 1.08 - 1.145$,
$R_d \approx (0.873-0.926) r_c$ and 
$\Delta R_d \approx 0.053 r_c$.
  The angle to the sonic point varies
from $\theta_*\approx 78^\circ$ (at
the lower limit of $\alpha$) to
$90^\circ$.
   The characteristic density in
this case is
$\rho_0 \approx 
10^{-12}{\rm g}/{\rm cm}^3$,
assuming that all of the disk accretion
goes into transonic flow along
the field lines.
   The ratio $r_c/R_{star}=8.6$
is much smaller than for the neutron
star case, and 
$v_{ff}(R_{star})
\approx 365~{\rm km/s}$.
   The angle at which the flow
hits the star is $\theta_{star}
\approx 20^\circ$.
  At this point the flow velocity
is at an angle $\approx 10.4^\circ$
with respect to $\hat{{\bf R}}$.

\section{Conclusions}

   This work analyzes the stationary 
ideal magnetohydrodynamic funnel flows
along the magnetic field lines of
a rotating star with an aligned dipole
magnetic field.  
   Isentropic flow is assumed so
that a given plasma blob maintains
$p/\rho^\gamma=$ const with
$1 < \gamma \leq 5/3$, and
the magnetic field is assumed 
strong, $\rho {\bf v}^2 \ll {\bf B}^2/4\pi$.
   Earlier, Li and Wilson (1999) considered the
Bernoulli equation assuming $p \propto \rho$.
  Close to the disk the flow is subsonic.
There are three possible cases:  
(1) stationary transonic
flow where a sonic point exists along
the dipole field line;  
(2) stationary subsonic flow;
and (3) no stationary flow exists.
  Over a range of $R_d \lesssim r_c$,
subsonic inflows are possible, where
$R_d$ is the radius at which the dipole
field line intersects the disk and
$r_c$ is the `corotation radius.'  
   For $R_d > \sim r_c$ there are no
stationary flows.

  The transonic flows, which 
become
free-fall inflow near the star, are
possible only for
a narrow range of $R_d \sim r_c$. 
   At the sonic point these flows have
an approximate balance of the
centrifugal and gravitational force
along the field line.
    The  flows have different
behaviors for $\gamma > 7/5$ and $< 7/5$.
   For $\gamma >7/5$, the Mach number 
${\cal M} \equiv |{\bf v}|/c_s$ 
initially increases
with $R$ decreasing from $R_d$, but for 
$R$ decreasing from 
$\approx 0.22 R_d$ (for $\gamma=5/3$)
the Mach number decreases.
   In the other case,
$\gamma < 7/5$, ${\cal M}$
increases monotonically 
with decreasing $R$.
    The variation of {\it both} $v_p$ and 
${\cal M}$ are clearly important 
to the nature of the standing shock near
the star and to the determination of
emission line profiles. 
  We discuss numerical values for
funnel flows to neutron stars and
young stellar objects.
   For the neutron star case we
argue that $\gamma$ is likely
to be less than $7/5$.  
   This combined with the large
value of $r_c/r_{star}$ implies
that the Mach number of the flow
approaching the star should be
very much larger than unity.
    On the other hand for flows
to a forming star we expect $\gamma >7/5$.
This combined with the not very
large values of $r_c/r_{star}$ implies
only moderately large
Mach numbers near the star. 
     The flow solutions
analyzed in this work are 
also of interest for comparison 
with MHD simulations.

\acknowledgments{
    This research was
supported in part by 
NASA grants NAG5-9047 and NAG5-9735 
and NSF grant AST-9986936.
M.M.R. received partial support
from NSF POWRE grant
AST-9973366.}

\begin{deluxetable}{lcccccccccc}
\footnotesize
\tablecaption{Sample  Parameters\tablenotemark{a}} 
\tablehead{
\colhead{$P$}          & 
\colhead{$R_{star}$}           &
\colhead{$M$}                  & 
\colhead{$\dot{M}$}            &
\colhead{$L_{acc}$}            & 
\colhead{$r_c$}                &
\colhead{$r_A$}                &
\colhead{$\mu$}                &
\colhead{$\Omega r_c$}         &
\colhead{$\gamma$}
 }
\startdata
$1$~s                                 & 
$10^6$cm                             &
$1M_\odot$                           &
$10^{-9}M_\odot$/yr                  &
$0.85\!\!\times \!\!10^{37}$erg/s     & 
$1.5\!\!\times \!\!10^8$cm           &
$3.7\!\!\times \!\!10^8$cm           &
$10^{30}$Gcm$^3$                     &
$9.4\!\!\times \!\!10^8$cm/s         &
1.3                                  &
\nl
$5$~d                                 &
$10^{11}$cm                          &
$1 M_\odot$                          & 
$10^{-7}M_\odot$/yr                  & 
$0.85\!\!\times \!\!10^{34}$erg/s    &
$8.6\!\!\times \!\!10^{11}$cm        & 
$9.6\!\!\times \!\!10^{11}$cm        &
$10^{37}$ Gcm$^3$    &
$125$~km/s                            &
$5/3$
\nl 
\enddata
\tablenotetext{a}{$P=2\pi/\Omega$ is the star's
rotation period; 
$~~R_{star}$ its radius;
$~~M$ its mass;  
$~~\dot{M}$ is the accretion rate;
$~~L=GM\dot{M}/r_{star}$ is the accretion luminosity;
$~~r_c=(GM/\Omega^2)^{1/3}$ is the `corotation radius';
$~~r_A =[\mu^4/(2GM\dot{M}^2))]^{1/7}$ 
is the Alfv\'en radius of
the disk (see Shapiro and Teukolsky 1983);
$~~\mu$ is the magnetic moment of the star;
$~~\Omega r_c$ is the azimuthal velocity of the
disk at the corotation radius; and
$\gamma$ is the usual adiabatic exponent.}
\end{deluxetable}

\end{document}